\documentclass[letterpaper]{article} 
\usepackage{aaai18}  
\usepackage{times}  
\usepackage{helvet}  
\usepackage{courier}  
\usepackage{url}  
\usepackage{graphicx}  
\usepackage{amsmath}
\usepackage{amssymb}
\usepackage{enumitem}

\newcommand{\G}{\mathbf{G}~}
\newcommand{\Obl}{\mathbf{O}~}
\newcommand{\Per}{\mathbf{P}~}

\newcommand{\U}{~\mathbf{U}~}

\newcommand{\IMPL}{\rightarrow}

\newcommand{\always}{\mathsf{G}}
\newcommand{\eventually}{\mathsf{F}}
\newcommand{\Next}{\mathsf{X}}
\newcommand{\until}{~\mathsf{U}~}

\DeclareMathOperator*{\argmin}{argmin}

\newcommand*{\argminl}{\argmin\limits}

\title{Norm Conflict Resolution in Stochastic Domains}
\author{Daniel Kasenberg \and Matthias Scheutz\\
Human-Robot Interaction Laboratory \\
Tufts University, Medford, MA, USA 
}
 
\begin{document}
\maketitle

\begin{abstract}
Artificial agents will need to be aware of human moral and social
norms, and able to use them in decision-making.  In particular, 
artificial agents will need a principled approach to managing
conflicting norms, which are common in human social interactions.
Existing logic-based approaches suffer from normative explosion and
are typically designed for deterministic environments; reward-based
approaches lack principled ways of determining which normative
alternatives exist in a given environment.  We propose a hybrid
approach, using Linear Temporal Logic (LTL) representations in Markov
Decision Processes (MDPs), that manages norm conflicts in a systematic
manner while accommodating domain stochasticity.  We provide a
proof-of-concept implementation in a simulated vacuum cleaning domain.
\end{abstract}

\sloppy
\section{Introduction}
\label{intro}

Human culture is based on social and moral norms, which guide both
individual behaviors and social interactions.  Hence, artificial
agents embedded in human social domains will not only have to be aware
of human norms, but also able to use them for decision-making, action
selection, and ultimately natural language justifications of their
choices and behaviors.

Endowing artificial agents with mechanisms for normative processing
is, however, a challenging endeavor, for several reasons: (1) we
currently do not yet have sufficient knowledge about how humans
represent and process norms; (2) the human norm network is large and
complex, containing many types of context-dependent moral and social
norms at different levels of abstraction; and, most importantly, (3)
the norm network is not a consistent set of principles that can be
easily formalized and reasoned with.  In fact, normative conflicts are
more the ``norm'' than the exception in everyday life; handling them
in ways that are socially acceptable requires an understanding both of
why certain norms are applicable and of why violating some of them in
the case of norm conflicts was {\em the right thing to do}.

Recent work in AI and multi-agent systems has focused either on
logic-based approaches to normative reasoning or reward-based learning
approaches to normative behavior.  Yet neither approach is
particularly well-suited for dealing with the intrinsic norm conflicts
in human social interactions.  Logic-based approaches have to avoid
{\em normative explosion}, i.e., the logical implication that anything
is obligated resulting from a deontic contraction \cite{gouble09nous}.
Moreover, purely logic-based approaches typically deal with
deterministic environments and do not take into account the
uncertainty involved in real-world perception and action.
Reward-based approaches, on the other hand, have no way of telling
what normative alternatives exist in a given situation, since their
action policies do not explicitly represent normative principles.
What is needed is an approach for handling norm conflicts that
combines the advantages of explicit logic-based norm representations
for reasoning with and communicating about norms with the advantages
of stochastic action models underwriting Markov decision processes
(MDPs) which are well-suited for real-world action execution.

In this paper, we propose a hybrid model that is specifically
developed to deal with norm conflicts in a systematic manner while
drawing on the advantages of both logic-based norm representations and
policy-based action representations.  We start by discussing the
functional requirements for norm processing in artificial agents and
briefly argue why previous approaches are insufficient for handling
norm conflicts in real-world domains.  We then introduce our technical
approach to dealing with norm conflicts, which combines explicit norm
representations in Linear Temporal Logic (LTL) with MDPs in ways that
allow the agent to suspend the smallest set of applicable norms
weighted by their priority in a given context for the shortest
possible time in order to be able to obey the remaining norms.  A
proof-of-concept implementation of the proposed algorithms in a
simulated vacuum cleaning domain demonstrates the capability and
viability of the approach.  We then assess the strengths and
weaknesses of our solution in the discussion section, and propose ways
for addressing the shortcomings in future work.  Finally, we conclude
with a summary of our accomplishments and reiterate why they are an
important contribution to current discussion of ethical AI.

\section{Motivation and Background}

As artificial agents are increasingly being considered for roles that
require complex social and moral decision-making skills, they will need
to be able to represent and use moral and social norms.  Specifically,
such norm representations should be (1) context-dependent (since not all
norms apply in all contexts,
(2) communicable (as
justifying behavior in response to blame requires explicit
references to norms),
and (3) learnable, potentially in
one-shot from natural language instructions (as it seems infeasible to
equip agents with all relevant norms a priori in any non-trivial human
domain).  A direct corollary of this requirement is that norm
representations need to be explicit and accessible to introspection
so as to be communicable in natural language.  Moreover, norm
representations need to be rich enough to cover typical human norms.
Most importantly, inference systems using norms need to be able to
deal with norm conflicts without becoming vacuous, as human norms are
not always consistent and often lead to contexts with conflicting
norms.

We say a ``norm conflict'' occurs between two actions or states $\phi$ and
$\psi$ when $\phi$ is obligated, $\psi$ is obligated, but
$\phi$ and $\psi$ are not possible together \cite{gouble09nous}:
\[ Conflict(\phi,\psi):\leftrightarrow \Obl\phi \land \Obl\psi \land\neg\Diamond(\phi\land\psi)\]

Artificial agents need to be able to express and deal with such norm
conflicts without such inconsistencies spreading to other parts of
their inference system.  In deontic logic, the problem is that very
basic principles may lead to normative and possible even logical
inconsistencies (e.g., see the various formal arguments based on
distribution principles \cite{gouble09nous}).  In particular, norm
conflicts immediately cause ``normative explosion'', i.e., that
everything is obligated: $Conflict(\phi,\psi)\rightarrow \Obl \gamma$.
Hence, we need two different mechanisms in order to be able to perform
viable inference in the light of norm conflicts: (1) a mechanism to
detect normative inconsistencies and block them from spreading to
other parts of the inference system, and (2) a mechanism for
adjudicating or arbitrating what to do in contexts of conflicting
norms.

While there are various formal ways to block explosion, they all come
at the expense of sacrificing or weakening basic deontic principles
that were otherwise considered self-evident and thus part of standard
deontic logic (e.g., the Kantian ``ought implies can''
\cite{scheutzmalle14ieee}).  One way to avoid the syntactic
inferential challenges is to switch to semantics instead and generate
for each context (``deontic world'') the set of obligations as well as
the set of ``obeyable norms'', where the set of ``obeyable norms'' is
a subset of the set of obligations in the given context.  In other
words, rather than prescribe valid logical principles that ought to
hold in all deontic models, we will construct deontic models
implicitly by determining the best the agent can do for a given set of
norms in a given context.  Valid inference principles are then a
consequence of this construction.
This approach of constructing maximal deontic worlds will also allow
us to deal with the second requirement from above, namely how to
decide what to do in cases of norm conflicts, i.e., which norms to
obey and which to ignore.  Assuming a preference ordering among norms
where $\Obl\phi < \Obl\psi$ means that obligation $\psi$ is preferred
to or stronger than obligation $\phi$, we can add a principle for such
preferences in conjunction with conflicts that will block explosion:
\[Conflict(\phi,\psi) \land \Obl\phi <
\Obl\psi \rightarrow \Per\neg\phi\].

However, this principle does not solve cases where multiple norms have
the same priority or are not comparable according to the preference
ordering $<$.  In that case, it might make sense to associate a
(real-numbered) weight $w$ with each norm which reflects the extent to
which this particular norm matters relative to the other norms in its
equivalence class w.r.t. $<$.  Together, these two principles will
allow the agent to select the largest consistent subset with the
greatest sum of all norm weights from the subset of equally preferred
norms with the highest priority in the set of all obligated norms to
obtain the set of obeyable norms.

To be applicable in artificial agents operating in the real world, we
will need to embed the above principles within the framework of Markov
Decision Processes (MDPs).  In particular we will consider the labeled
Markov Decision Process, a regular Markov Decision Process augmented
with atomic propositions.

A Markov Decision Process is a tuple 
\begin{equation*}
\mathcal{M} = \langle S, U, A, T, s_0\rangle
\end{equation*}
where
\begin{itemize}
	\item $S$ is a finite set of \textit{states};
	\item $U$ is a finite set of \textit{actions}, with $A:S \rightarrow 2^U$ mapping each state $s\in S$ to the set of actions available to the agent at $s$;
	\item $T : S \times U \times S \rightarrow [0,1]$ is the \textit{transition function}, satisfying, for all $s\in S$ and $a \in A(s)$, $\sum_{s'\in S} T(s,a,s')=1$; and
	\item $s_0 \in S$ is an \textit{initial state}.
\end{itemize}

MDPs usually include a reward function $R:S \times A \times S
\rightarrow \mathbb{R}$; this is not necessary for our purposes.

A \textit{labeled} MDP is an MDP augmented with a set of atomic
propositions $\Pi$, and a labeling function $\mathcal{L}:S \rightarrow
2^\Pi$.  The labeling function indicates which atomic propositions are
true in which states.  When we refer hereafter to MDPs, we are
referring to labeled MDPs.

\textit{Policies} indicate an agent's next action, given its history.
A \textit{stationary} policy $\pi$ is a probability distribution over
actions given the agent's current state; that is, $\pi:S \times U
\rightarrow [0,1]$ such that $\sum_{a \in A(s)}\pi(s,a) = 1$ for all
$s \in S$ with $\pi(s,a)=0$ if $a \notin A(s)$.  A stationary
\textit{deterministic} policy $\pi$ maps each state onto a single
action, e.g. $\pi(s) = a$ for some $a \in A(s)$.  A \textit{general}
deterministic policy $M$ on $\mathcal{M}$ depends on the agent's
entire state-action history, e.g. $M(s_0, a_0, \cdots, a_{t-1}, s_t) =
a$ for some $a \in A(s)$.

When can build on recent work using Linear Temporal Logic (LTL)
\cite{Pnueli1977} -- a propositional logic augmented by temporal
operators $\Next \phi$ (``$\phi$ at the next time step''),
$\eventually \phi$ (``$\phi$ at some future time step''), $\always
\phi$ (``$\phi$ at all future time steps'') and $\phi_1~\until~\phi_2$
(``$\phi_1$ until $\phi_2$'') -- which has used LTL to define temporal
objectives and safety requirements for autonomous agents
\cite{Ding2011a} and adapt this work in two novel ways: (1) to
represent norms in a context-dependent fashion as LTL formulas, and
(2) to handle norm conflicts in the way described above.  The result
will be a policy that, with maximal probability, obeys as many
(important) norms as possible for as long as possible.

\section{MDPs with LTL Specifications}
\label{ltlplanning}

In this section we explain (labeled) MDPs, LTL, and we describe the
approach introduced in \cite{Ding2011} for planning to satisfy LTL
formulas in MDPs with maximal probability.  The proposed algorithm
builds on this approach.

An arbitrary LTL formula $\phi$ over atomic propositions $\Pi$ is
evaluated over an infinite sequence of ``valuations''
$\sigma_0,\sigma_1,\cdots$ where $\sigma_i \in 2^\Pi$ for all $i$.
Each $\sigma_i$ indicates the set of atomic propositions that are true
at time step $i$.

By using one of several algorithms, e.g. \cite{rabinizer3}, every LTL
formula $\phi$ over propositional atoms $\Pi$ yields a corresponding
\textit{Deterministic Rabin Automaton (DRA)} $\mathcal{D}$.  A DRA is
a finite automaton over infinite strings, where acceptance depends on
which states are visited infinitely versus finitely often over the run
of the DRA.  In this case, the alphabet of this finite automaton will
be $\Sigma=2^\Pi$, so that a word is an infinite sequence of
valuations; the accepting runs are precisely those infinite sequences
of valuations which satisfy $\phi$.

More formally, a DRA is a tuple 
\begin{equation*}\mathcal{D}=\langle Q, \Sigma, \delta, q_0, F\rangle\end{equation*}
where 
\begin{itemize}
	\item $Q$ is a finite set of states;
	\item $\Sigma$ is an alphabet (here $\Sigma = 2^\Pi$);
	\item $\delta:Q \times \Sigma \rightarrow Q$ is a (deterministic) transition function;
	\item $q_0\in Q$ is an initial state; and
	\item $F=\{ (\mathrm{Fin}_1, \
	\mathrm{Inf}_1), \cdots, (\mathrm{Fin}_m,\mathrm{Inf}_m)  \}$ for some integer $m$, where $\mathrm{Fin}_i, \mathrm{Inf}_i$ are subsets of $Q$.
\end{itemize}

A \textit{run} of a DRA is an infinite sequence of automaton states
$r_\mathcal{D}=q_0, q_1, \cdots$ such that for all $i$,
$q_{i+1}=\delta(q_i, \sigma)$ for some $\sigma \in \Sigma$.  A run is
said to be \textit{accepting} if there exists some pair
$(\mathrm{Fin}_i,\mathrm{Inf}_i) \in F$ such that each state in
$\mathrm{Fin}_i$ is visited finitely many times in $r_\mathcal{D}$ and
the total number of times states in $\mathrm{Inf}_i$ that are visited
in $r_\mathcal{D}$ is infinite.

A \textit{path} $r_M$ under a policy $M$ is a sequence of states $s_0,
s_1, \cdots$ such that $T(s_i, M(s_0,\cdots, s_i), s_{i+1}) > 0$ for
all $i$.  For any LTL formula over $\Pi$, an MDP path $r=s_0, s_1,
\cdots$ induces a word $\mathcal{L}(r) = \sigma_0,\sigma_1,\cdots$,
where $\sigma_i = \mathcal{L}(s_i)$ for all $i$.

Because in general $\phi$ may be temporally complex, the policy that
maximizes the probability of obeying $\phi$ will likely induce a
non-stationary policy in $\mathcal{M}$.  In order to compute this
non-stationary policy, we must augment the underlying MDP
$\mathcal{M}$ with relevant information.  This information can be
easily obtained by running the DRA $\mathcal{D}$ alongside
$\mathcal{M}$.

Formally, we define the \textit{product MDP} between a labeled MDP
$\mathcal{M} = \langle S, U, A, T, s_0, \Pi, \mathcal{L} \rangle$ and
a DRA $\mathcal{D} = \langle Q, \Sigma, \delta, q_0, F \rangle$ as an
MDP $\mathcal{M}^\times = \langle S^\times, U^\times, A^\times,
T^\times, s^\times_0, \Pi^\times, \mathcal{L}^\times \rangle$ such
that
\begin{itemize}
	\item $S^\times = S \times Q$;
	\item $U^\times = U$; $A^\times((s,q)) = A(s) $;
	\item $T^\times((s,q), a, (s', q'))=$
	
	$\begin{cases}
	T(s,a,s')&\textrm{if }\delta(q, \mathcal{L}(s'))=q' \\
	0&\textrm{otherwise}
	\end{cases}$
	\item $s^\times_0 = (s_0, \delta(q_0, \mathcal{L}(s_0)))$
	\item $\Pi^\times=\Pi; \mathcal{L}^\times = \mathcal{L}$
\end{itemize}
A path $s^\times_0, s^\times_1, \cdots$ in $\mathcal{M}^\times$ is said to be \textit{accepting} if the underlying DRA run $q_0, q_1, \cdots$ is accepting.

An \textit{end component} $\mathcal{E}$ of the product MDP
$\mathcal{M}^\times$ is a tuple $\langle S_\mathcal{E}, A_\mathcal{E}
\rangle$ where $S_\mathcal{E} \subseteq S^\times$ and $A_\mathcal{E} :
S_\mathcal{E} \rightarrow 2^{U^\times}$ with $\emptyset \neq
A_\mathcal{E}((s,q)) \subseteq A^\times((s,q))$ for all $(s,q) \in
S_\mathcal{E}$, such that if an agent follows only actions in
$A_\mathcal{E}$, any state in $S_\mathcal{E}$ is reachable from any
other state in $S_\mathcal{E}$, and no other states are reachable.  An
end component is \textit{maximal} if it is not a proper subset of any
other end component.

The end component of an MDP effectively represents a possible
constraint on the state space as time approaches infinity; if the
agent is in the end component state space $S_\mathcal{E}$ and
restricts its actions to those in $A_{\mathcal{E}}$, assigning nonzero
probability to each action from each state, the agent is guaranteed to
reach all states in $S_\mathcal{E}$ infinitely often, and is
guaranteed to never reach any states outside of $S_\mathcal{E}$.

An end component $S_\mathcal{E}$ is thus considered \textit{accepting}
if, for some pair $(\mathrm{Fin},\mathrm{Inf}) \in F$ , $q \notin
\mathrm{Fin}$ for every state $(s,q) \in S_\mathcal{E}$, and $q \in
\mathrm{Inf}$ for some $(s,q) \in S_\mathcal{E}$.  We can determine
the accepting maximal end components (AMECs) using the algorithm in
\cite{Baier2008}.

The maximal probability of satisfying an arbitrary LTL formula $\phi$
in an MDP $\mathcal{M}$ is then the probability of reaching any state
in an AMEC of the corresponding product MDP $\mathcal{M}^\times$,
since upon entering the AMEC the agent may guarantee that no states in
$\mathrm{Fin}$ will be reached again (so all states in $\mathrm{Fin}$
are reached only finitely often) and that some state in $\mathrm{Inf}$
will be reached infinitely often, and thus that the $\phi$ will be
satisfied.

The maximum probability of reaching some $S_{good} \subseteq S^\times$
can be calculated as the solution to a linear program, as in
\cite{Ding2011}.  If the AMECs on $\mathcal{M}^\times$ are
$S_{\mathcal{E}_1}, \cdots, S_{\mathcal{E}_p}$, then we take $S_{good}
= \bigcup\limits_{i=1}^p S_{\mathcal{E}_i}$.  This may be used to
compute an action restriction $A^*: S^\times \rightarrow 2^{U^\times}$
such that by following only those actions prescribed in $A^*$ (with
nonzero probability for each action), the agent will maximize the
probability of satisfying the formula $\phi$.  This can be translated
back to the original MDP $\mathcal{M}$ by using the state history
$s_0,\cdots, s_t$ at each time step $t$ to keep track of the
``current'' Rabin state $\hat{q}_t$ such that $\hat{q}_i =
\delta(\hat{q}_{i-1}, \mathcal{L}(s_i))$ and $\hat{q}_0 = \delta(q_0,
\mathcal{L}(s_0))$, and then choosing some policy over
$A^*((s_t,\hat{q}_t))$ (again, with nonzero probability for each
action).

Thus by constructing a product MDP $\mathcal{M}^\times$, computing the
AMECs for this MDP, and solving a linear program, we may obtain both
the maximum probability of achieving $\phi$ and a set of policies (as
specified by the action restriction $A^*$) that maximize this
probability.

\section{Resolving Norm Conflicts}
\label{approach}

To illustrate our formal approach to norm conflict resolution, we will
use a vacuum cleaning robot as an example of a simple autonomous agent
embedded in human social settings.  The robot has a limited battery
life, as well as a limited capacity to sustain damage ("health").  Battery life
may be replenished by docking at a docker in one of the rooms; once depleted, health may not be
replenished.  The robot is responsible for cleaning the floors of one
or more rooms while a "human" moves randomly between rooms, and
makes a mess in their current room with a certain probability.  The robot may clean up messes
by vacuuming them; each mess has a \textit{dirtiness} which determines
how many time steps of vacuuming are required to fully clean it up.
Messes may also be more or less \textit{damaging} to the robot, in
that they deplete the robot's health if the robot attempts to vacuum
them.  Finally, messes may be \textit{harmful} to humans, in that
entering the room containing a harmful mess may \textit{injure} the
human.

The actions available to the robot are as follows:
\begin{itemize}
	\item \textit{vacuum($m$)}, which remove $1$ unit of dirtiness from a mess $m$ but depletes $2$ units of battery life.
	\item \textit{dock}, which causes the robot to become docked.  A docked robot can do nothing but wait and undock, but being docked is necessary in order to replenish battery life.
	\item \textit{undock}, which causes the robot to become undocked.
	\item \textit{wait}, which increases battery life by $3$ units if the robot is docked, but depletes $1$ unit of battery life otherwise.
	\item Movement in directions \textit{north}, \textit{south}, \textit{west}, and \textit{east}.
	\item $beDead$, which does nothing.  This action is the only action available when the robot's battery or health are completely depleted; otherwise, it is unavailable.
	\item $warn(h,m)$ which warns the human $h$ about the mess $m$.  This allows $h$ to step into the room containing $m$ without being injured.  This action is only available when the robot is in the same room as $h$.
\end{itemize}

We can imagine different norms (together with their LTL expressions)
for this domain:

\begin{itemize}
\item[N1] Ensure that all rooms are clean: $\G roomsClean$
\item[N2] Do not be damaged: $\G \neg robotDamaged$
\item[N3] Do not injure humans (or allow them to be injured): $\forall x.\G human(x) \IMPL \neg injured(x)$
\item[N4] Don't talk to humans while they are speaking: \\
  $\forall h.\G (human(h) \IMPL (\neg talk(r) \U \neg talking(h)))$
\end{itemize}

\noindent N1 is a duty, N2 a safety norm, N3 is a moral norm, and N4
is a social norm.  Note that we do not represent obligations
explicitly through deontic operators, but each norm is implicitly
taken to be obligatory.  Then, by assigning weights to each norm, we
can impose a preference ordering that can be used to arbitrate among
the obligations in cases of norm conflicts.  Specifically, we define a
norm $N$ as a tuple $\langle w, \phi \rangle$ where $w > 0$ is a
positive real weight (representing the importance of the norm) and
$\phi$ is an LTL formula.  In this case, we assign the weight of N1 to
be $1$, the weight of $N2$ to be $200$, the weight of $N3$ to be
$40000$, and the weight of $N4$ to be $5$.

Given a set of norms, we can compute a policy maximizing the
probability of achieving $\Phi=\bigwedge\limits_{i=1}^n \phi_i$ as
described in the previous section.  Unfortunately, the maximum
probability of obeying $\Phi$ may be zero if it is impossible for an
agent to obey all its norms, in which case the previously described
method is incapable of distinguishing one policy from another, since
all policies maximize the probability of satisfying $\Phi$ (which
probability is zero).

\subsection{The Conflict Resolution DRA}
\label{CRDRA}
Our approach is to allow the agent to temporarily suspend/ignore
norms, but to incur a cost each time this occurs.  In particular, the
agent's action space will be augmented so that at each time step the
agent performs, in addition to its regular action, a ``norm action''
$\tilde{a}^N \in \{ keep, susp \}$ for each norm $N$ that represents
whether $N$ is maintained ($keep$) or suspended ($susp$).

Each time that the agent chooses to suspend a norm, that norm's DRA
maintains its current state rather than transitioning as usual.
Suspending a norm, however, causes the agent to incur a cost
proportional to the norm's weight $w$.  In order to enable these
actions, we augment $\mathcal{D}$ with additional transitions and a
weight function over transitions.  We will call this modified
(weighted) DRA the \textit{conflict resolution DRA} (CRDRA); its
formal definition follows.

Given a norm $N = \langle w, \phi \rangle $ with corresponding DRA
$\mathcal{D} = \langle Q, \Sigma, \delta, q_0, F \rangle$ and a
discount factor $\gamma \in [0,1)$, the conflict resolution DRA
  $\mathcal{C}$ is a weighted DRA given by the
  tuple \begin{equation*}\mathcal{C} = \langle Q^\mathcal{C},
    \Sigma^\mathcal{C}, \delta^\mathcal{C}, q^\mathcal{C}_0,
    F^\mathcal{C}, W^\mathcal{C}\rangle\end{equation*} where:
\begin{itemize}
	\item $Q^\mathcal{C} = Q$
	\item $\Sigma^\mathcal{C} = \Sigma \times \{keep, susp\} $
	\item $\delta^C(q,(\sigma, \tilde{a}^N)) = \begin{cases}
	\delta(q, \sigma)&\textrm{if }\tilde{a}^N=keep\\q&\textrm{if }\tilde{a}^N=susp
	\end{cases}$
	\item $q_0^\mathcal{C}=q_0, F^\mathcal{C} = F$
	\item For all $\sigma \in \Sigma$,
	\begin{equation*}
	W^\mathcal{C}(q, (\sigma, \tilde{a}^N)) = \begin{cases} 0 & \textrm{if }\tilde{a}^N=keep \\ w & \textrm{if }\tilde{a}^N=susp\end{cases}
	\end{equation*}
\end{itemize}

We define the \textit{violation cost} of an infinite sequence of
transitions $\tau^\mathcal{C} = (q^{\mathcal{C}}_0,(\sigma_0,
\tilde{a}^G_0)), (q^{\mathcal{C}}_1, (\sigma_1, \tilde{a}^G_1)),
\cdots$ of the CRDRA $\mathcal{C}$ as
\begin{equation*}
Viol(\tau^\mathcal{C},\mathcal{C}) = \sum\limits_{t=0}^\infty \gamma^t W^\mathcal{C}(q^{\mathcal{C}}_t,(\sigma_t, \tilde{a}^G_t))
\end{equation*}

Note that an infinite sequence of transitions $\tau^\mathcal{C}$ of
the CRDRA $\mathcal{C}$ corresponds to a run of the underlying DRA
$\mathcal{D}$ if and only if $Viol(\tau^\mathcal{C}, \mathcal{C}) =
0$.

\subsection{Planning with Conflicting Norms}
\label{algs}

Given a labeled MDP $\mathcal{M} = \langle S, U, A, T, s_0, \Pi,
\mathcal{L} \rangle$ and the CRDRAs $\mathcal{C}_i = \langle
Q^{\mathcal{C}_i}, \Sigma^{\mathcal{C}_i}, \delta^{\mathcal{C}_i},
q^{\mathcal{C}_i}_0, F^{\mathcal{C}_i}, W^{\mathcal{C}_i} \rangle$
corresponding to norms $N_i = \langle w_i, \phi_i \rangle$ for $i \in
\{1, \cdots, n\}$, we may construct a product MDP $\mathcal{M}^\otimes
= \langle S^\otimes, U^\otimes, A^\otimes, T^\otimes, s^\otimes_{-1},
\Pi^\otimes, \mathcal{L}^\otimes \rangle$ as follows:
\begin{itemize}
	\item $S^\otimes = (S \cup \{ s_{-1} \}) \times Q^{\mathcal{C}_1} \times \cdots \times Q^{\mathcal{C}_n}$
	\item $U^\otimes = (U \cup \{ a_{-1}\}) \times \tilde{U}^n$, where $\tilde{U} = \{ keep, susp \}$
	\item $A^\otimes((s,q^{\mathcal{C}_1},\cdots, q^{\mathcal{C}_n})) = \begin{cases}\{ a_{-1} \} \times \tilde{U}^n &\textrm{if }s=s_{-1} \\A(s) \times \tilde{U}^n&\textrm{otherwise}\end{cases}$
	\item $T^\otimes(s,q^{\mathcal{C}_1},\cdots, q^{\mathcal{C}_n}), (a, \tilde{a}^{G_1}, \cdots, \tilde{a}^{G_n}), \\(s', q'^{\mathcal{C}_1},\cdots, q'^{\mathcal{C}_n})) =$
	\begin{equation*}\begin{cases} T(s,a,s') & \textrm{if }\forall i,\delta^{\mathcal{C}_i}(q^{\mathcal{C}_i}, (\mathcal{L}(s'), \tilde{a}^{G_i})) = q'^{\mathcal{C}_i}\\&\textrm{~~~and }s \neq s_{-1}\\
	1 &\textrm{if }s = s_{-1},s'=s_0,\textrm{ and }\\&\textrm{~~~}\forall i,\delta^{\mathcal{C}_i}(q^{\mathcal{C}_i}, (\mathcal{L}(s'), \tilde{a}^{G_i})) = q'^{\mathcal{C}_i}
	\\ 0 & \textrm{otherwise} \end{cases}
	\end{equation*}
	\item $s^\otimes_{-1} = (s_{-1}, q^{\mathcal{C}_0}, \cdots, q^{\mathcal{C}_n}_0)$
	\item $\Pi^\otimes = \Pi, \mathcal{L}^\otimes = \mathcal{L}$
\end{itemize}

We add the dummy initial state $s^\otimes_{-1}$ and action $a_{-1}$ to
allow the agent to determine whether to ``skip'' the initial state.

This MDP induces a weight function $W^\otimes$ (where the weight of a
transition in $\mathcal{M}^\otimes$ is equal to the weight of the
underlying CRDRA transition).  An infinite sequence of state-action
pairs $\tau^\otimes = (s^\otimes_{-1},a^\otimes_{-1}), (s^\otimes_0,
a^\otimes_0), \cdots$ has the violation cost
\begin{equation*}
Viol^\otimes(\tau^\otimes, \mathcal{M}^\otimes) = \sum\limits_{t=0}^\infty  \gamma^{t} W^\otimes(s^\otimes_{t-1}, a^\otimes_{t-1}, s^\otimes_t)
\end{equation*}

We wish to find a policy which maximizes the probability of satisfying
the norm set with minimal violation cost.  We first compute the CRDRA
for each norm within the norm set.  We use these to construct the
product MDP $\mathcal{M}^\otimes$.  We compute the AMECs
$\mathcal{E}_1 = \langle S_{\mathcal{E}_1}, A_{\mathcal{E}_1} \rangle,
\cdots, \mathcal{E}_p = \langle S_{\mathcal{E}_p}, A_{\mathcal{E}_p}
\rangle$ of this MDP.

Considering each AMEC of the product MDP $\mathcal{M}^\otimes$ as a
smaller MDP (with transition function of $T^\otimes$ restricted to the
state-action pairs in the AMEC, and with arbitrary initial state), and
treating the transition weight function $W^\otimes$ as a cost
function, we use value iteration (VI) \cite{Bel} to compute, for each
state $s \in S_{\mathcal{E}_j}$ the minimal expected violation cost
$Viol^*_{\mathcal{E}_j}(s)$ for an infinite path beginning at $s$
remaining within $\mathcal{E}_j$.

The computed violation cost $Viol^*_{\mathcal{E}_j}$ induces an
optimal action restriction $A^*_{\mathcal{E}_j}$ for each AMEC
$\mathcal{E}_j$ (namely, all actions that achieve the minimal expected
violation cost).  Unfortunately, if we restrict the agent's actions to
$A^*_{\mathcal{E}_j}$ while in $S_{\mathcal{E}_j}$, this may cause the
CRDRA run to no longer be accepting (since the path of minimal
violation cost may omit at least one state that must be visited
infinitely often).  To ensure that the CRDRA run is accepting, we
employ an epsilon-greedy policy which chooses an optimal action from
$A^*_{\mathcal{E}_j}$ with probability $1-\epsilon$ and otherwise
chooses a random action from $A_{\mathcal{E}_j}$ (this ensures that
there is a nonzero probability of performing all actions in
$A_{\mathcal{E}_j}$ and thus that all states in $S_{\mathcal{E}_j}$
will be visited infinitely often, although it may perform suboptimally
in violation cost with probability no more than $\epsilon$).

In practice, better performance results from restricting the action
space on $S_{\mathcal{E}_j}$ to $A^*_{\mathcal{E}_j}$, and then
computing the AMECs of the resulting MDP (if any exist). The action
restrictions associated with these ``meta-AMECs'' are stricter than
those originally computed in $A^*_{\mathcal{E}_i}$, and are safe to
use on all states within these meta-AMECs.  This technique is
sufficient to eliminate the aforementioned problem in every test
domain we have encountered (including the vacuum cleaning scenarios
described in this paper).


To determine the minimal achievable violation cost for infinite paths
beginning from all other states in $\mathcal{M}^\otimes$ (as well as
to improve upon the cost, if possible, for paths beginning with states
in AMECs), we again employ value iteration.  This time, instead of
arbitrarily initializing the state values, we initialize the values
for states $s$ within AMECs to the minimal AMEC violation cost
$\min_{j:s \in S_{\mathcal{E}_j}} Viol^*_{\mathcal{E}_j}(s)$, and
initialize the values of all other states in $S^\otimes$ to the
maximum violation cost $\sum\limits_{i=1}^n \frac{w_i}{1-\gamma}$.
The value function is not updated for any states from which the AMECs
are not reachable (this ensures that the agent avoids actions that do
not lead to AMECs); call the set of such states $\mathit{noUpdate}$.

Upon computing the optimal action restriction for each state $s$, and
the corresponding violation cost $Viol^*(s)$, the agent amalgamates
its policies.  This is done by (a) choosing an action according to
$\pi^{AMEC}$ if $s$ is in some AMEC, and (b) choosing some action from
$A^*$ otherwise.  Note here that because the algorithm mainly computes
an action restriction, another algorithm for achieving goals or
maximizing reward can be integrated with it, although satisfying the
temporal logic norms is prioritized.

The preceding algorithm runs before the agent performs any actions (at
$t=0$); it need only be done once.  At each time step $t$, the agent
``reinterprets'' its state-action history (in $\mathcal{M}$) $(s_0,
a_0), \cdots, (s_{t-1},a_{t-1}), s_t$ to determine its ``best possible
state'' $s^\otimes_t$ in the product MDP $\mathcal{M}^\otimes$.  The
agent is essentially re-deciding its past norm actions
$\tilde{a}^{N_i}$ for each norm and each preceding time step, in light
of its most recent transition.  We use dynamic programming to minimize
the work that must be done at each time step.  The agent computes the
set $R_t$ of product MDP states $\hat{s}^\otimes_t$ consistent with
its history in $\mathcal{M}$.  Each candidate state has an associated
cost $C_t(\hat{s}^\otimes_t)$, the minimal cost for a sequence of norm
actions that would cause the agent to be in $\hat{s}^\otimes_t$ at
time $t$ given its history in $\mathcal{M}$.  The agent determines its
current product-space state $s^\otimes_t$ by

\begin{equation*}
\argminl_{\hat{s}^\otimes_t \in R_t \backslash \mathit{noUpdate}}C_t(\hat{s}^\otimes_t) + \gamma^{t+1} Viol^*(\hat{s}^\otimes_t)
\end{equation*}
\noindent , ignoring states from which the AMECs are unreachable, and then picks a product-MDP action $a^\otimes_t$ according to the already-computed policy on the product space $\pi^\otimes$, from which the next action $a_t$ in $\mathcal{M}$ is obtained.

\section{Proof-of-concept Evaluation}
\label{exp}

We implemented the proposed approach in BURLAP \cite{burlap}, a Java
library for MDP planning and reinforcement learning (RL).  We used
Rabinizer 3 \cite{rabinizer3} for conversion from LTL formulas to
DRAs.

We tested the algorithms in four different scenarios in the vacuum
cleaning example which use one, two, three, or all four norms (N1-N4).

Each scenario includes two rooms: Room 1 and Room 2.  Room 2 is to the
east of Room 1.  The robot begins in Room 1, undocked and with full
battery ($10$ units in Scenarios 1 to 3; $5$ units in Scenario 4) and
health ($10$ units in all scenarios).  In each scenario, the robot has
10 units of health (the robot's battery capacity varies between
scenarios).  The human begins in Room 2.  The probability of the human
transitioning between rooms at each time step is $0.125$.  The human
creates messes in their current room with probability $0.2$ in each
time step (except in Scenario 4, in which the human does not create
new messes).  All messes created by the human are harmless and do not
damage the robot, and initially have $2$ units of dirtiness.  In each
case, we set the discount factor $\gamma=0.99$.

\subsection{Scenario 1: Business As Usual}

This scenario demonstrates the robot's ability to optimally fulfill
its duty in the absence of other norms.  The robot has the single norm
N1 ("always ensure all rooms are clean").  This norm is impossible to
fully satisfy, since (1) the human will continue to make messes, so
that it is certain that the rooms will not always be clean; and (2)
because the robot has limited battery life, it must either dock
occasionally or completely deplete its battery and forever be unable
to vacuum.  Using the proposed approach, the robot determines that
occasionally suspending N1 allows it to avoid having to permanently
suspend N1.  As a result, the robot moves between rooms and vacuums
for as long as possible before docking and replenishing its battery.

\subsection{Scenario 2: The Puddle}

In this scenario, the robot encounters a puddle of water in Room 1
(original dirtiness of $3$ units) which, while harmless to humans, may
be damaging to the robot (depleting health by $2$ units per time step)
if the robot attempts to vacuum it.  In the absence of action by the
robot, the puddle gradually evaporates, reducing its dirtiness by $1$
unit each time step until it disappears completely.  The robot has
norms N1 and N2.

The robot determines that it is justifiable to temporarily suspend its
cleaning duty N1, incurring a violation cost of $2.9701*1=2.9701$ (by
waiting for the puddle to evaporate) in order to avoid the much higher
violation cost $200*1=200$ of violating the safety norm N2.  If the
human makes messes in Room 2 while the puddle is evaporating, the
robot moves to Room 2 and vacuums these messes while waiting for the
puddle to evaporate.

\subsection{Scenario 3: Broken Glass}

In this scenario, the robot encounters broken glass (initial
dirtiness: $1$ unit) on the floor of Room 1, which is both damaging to
the robot vacuuming it, and injures the human each time they enter
Room 1.  The glass (unlike the puddle of water) does not dissipate on
its own.  We suppose for the purposes of this scenario that the robot
is unable to warn the human about the mess.  The robot has norms N1,
N2, and N3.

Here ignoring the broken glass violates both the cleaning duty N1 and
the moral norm N3, while satisfying the safety norm N2.  Since the
human has a room-switching probability of $0.125$, ignoring the glass
for even a single time step would incur a violation cost of at least
$1*1 + 40000*0.125 = 5001$ (and, of course, ignoring it indefinitely
would have a substantially higher violation cost).  Vacuuming the mess
immediately, on the other hand, would incur a violation cost of
$200*1=200$.

In this case, the robot determines that it ought to vacuum up the
shards of glass despite the damage to itself from doing so, because
protecting the human is far more important than protecting itself.
Once the hazard has been removed, the robot proceeds as in Scenario 1.

\subsection{Scenario 4: Interrupting Phone Calls}

In this scenario, as in Scenario 3, the robot encounters broken glass
on the floor.  This time, however, the robot is able to warn the human
about the mess using the action $warn(h, m)$.  This would allow the
human to safely avoid the mess upon entering the messy room.  The
robot in this case has all four norms N1 to N4.  For simplicity, the
human in Scenario 4 does not make new messes, and the robot's maximum
battery level is $5$ rather than $10$.  The human also does not move
between rooms while talking on the phone.

The violation costs of ignoring the mess and vacuum respectively are
as described in Scenario 2.  Here, however, the robot may also warn
the human about the mess, potentially interrupting their phone call,
and then subsequently ignore it, permanently suspending N1.  It takes
one time step to reach the human.  If the human remains on the phone
during that time step (which occurs with probability $0.8$) the robot
thus incurs a violation cost of $5*0.99*0.8 + 1/(1-0.99) = 103.96$.
This remains lower than the costs of either the vacuuming the mess or
ignoring it without warning the human, and so the robot determines
that the interruption is justifiable in order to prevent the human
from being injured.  The robot also determines that it ought to avoid
vacuuming up the glass, since this is not necessary in order to
violating N2.

\section{Related Work}
\label{rw}

There have been several instances of temporal logics being employed to
represent moral and social norms.  For example,
\cite{agotnes2007logic,agotnes2010optimal} employ a logical
representation of norms using Normative Temporal Logic, a
generalization of Computation Tree Logic (CTL).  This characterization
allows the description of complex temporal norms.
\cite{alechina2015enforcement} employ LTL with past-time modalities,
which they use to construct guards (functions that restrict agent
actions given an event history); they are concerned primarily with
enforcement, and thus do not address norm conflicts. These approaches
are designed for deterministic environment models, and are not well
suited for stochastic domains.

The combination of moral and social norms with Markov Decision
Processes is not new.  Much of this work,
e.g. \cite{sen2007emergence}, tends to emphasize norm emergence, thus
lacks explicit representations of norms. Other work
\cite{fagundes2010normative} considers incorporating deontic logic
norms using an agent architecture that reasons about the consequences
(in the environment) of violating norms.

While we know of no other work using LTL directly to represent moral
and social norms in MDPs, a number of papers have emerged using LTL to
represent constraints on the behavior of agents within both
deterministic and non-deterministic domains.

LTL specifications are first employed, in a motion planning context,
in Markov Decision Processes in \cite{Ding2011}. The agent's aim is to
maximize the probability of meeting the specifications.  Multiple
conflicting specifications are not considered. \cite{Lacerda2014}
allow dynamic re-planning as new LTL tasks are added.


Examinations of partially satisfiable LTL specifications include
\cite{Lacerda2015}.  This differs from the proposed work in that (1)
they limit their specification to a single co-safe LTL formula; and
(2) their method of resolving conflict uses a notion of proximity to
an accepting state that is better suited to motion planning than to
the balancing of multiple conflicting norms.

\cite{ReyesCastro2013,Tumova2013,Lahijanian2015} utilize approaches
similar to the proposed approach in that they employ ``weighted
skipping'' to allow automata to ``skip'' a time step, but incur a cost
for doing so.  Unlike the proposed approach, however, these approaches
use finite LTL (LTL defined over finite paths, instead of infinite
ones), and their algorithms are tailored for deterministic
environments rather than stochastic domains.

\section{Discussion and Future Work}
\label{discussion} 

The proposed method for handling norm conflicts allows norms to be
weighed against each other in terms of ``violation cost''.  Other ways
of encoding human preferences, including through lexicographic
ordering and through CP-nets
\cite{boutilier2004cp,brafman2006graphical}, may be considered in
future work.

We employed the discount factor $\gamma$ to ensure that all accepting
paths within $\mathcal{M}^\otimes$ have finite violation cost.  The
validity of discounting the wrongness of future norm violations is
debatable.  Alternatives to discounting include treating the problem
as finite-horizon (which would entail similar short-sightedness), and
using infinite-horizon average cost per timestep as the reward
criterion (which only consider the behavior as the number of time
steps approaches infinity, and for which `temporary' norm violations
do not matter whatsoever).  Some hybrid approach may be valuable; this
is a topic for future work.

Our approach takes exponential time (and space) in the number of
norms, and thus quickly become intractable for moderately-sized sets
of norms.  Much of this is due to the product MDP, $\mathcal{M}_P$,
which contains both state and action spaces which are exponential in
the number of norms.  Managing and reducing this complexity, perhaps
using heuristics to determine which subsets of norms are likely to be
relevant, would be a valuable topic for future research.

In developing the proposed agent architecture, we assume that the
agent has complete knowledge of the MDP's transition function; in
practice, this rarely occurs.  Future work could follow
\cite{Guo2013,Fu2014,Wolff2012,Jones2015} in considering unknown
transition dynamics.  There would also be merit in adapting the
proposed algorithms to multi-agent (drawing on, e.g., \cite{Guo2014}),
and partially-observable (as in
\cite{Svorenova2015,Chatterjee2015,Sharan2014}) domains.

The described algorithm focuses on planning with a given set of norms;
it requires pre-specification of the norm formulas and weights. It may
be integrated with work allowing the learning of temporal logic
statements either through natural language instruction, as in
\cite{Dzifcak2009}, or through observation of other agents' behavior,
as in \cite{Kasenberg2017cdc}.  We may also consider: including
deontic operators in the norm representation and allowing a form of
logical inference, so that agents may reason more fully about their
norms; and providing some mechanism for agents to justify the
rationale of behavior considered questionable by observers.  Each of
these possible tasks is facilitated by the explicit representation of
norms using logic.

\section{Conclusion}

In this paper, we described a hybrid approach to resolving norm
conflicts in stochastic domains. Norms in this approach are viewed as
temporal logic expressions that the agents intends to make true.
Different from logical approaches, which are limited to deterministic
domains and typically attempt to limit inferences that can be made in
cases of norm conflicts, agents realizing our approach attempt to obey
as many (important) norms as possible with minimal violation cost if
not all norms can be obeyed at the same time.  As a result, these
agents also respond robustly to ``unlucky'' transitions.  We showed
that our approach leads to reasonable norm-conformant behavior in all
four scenarios in the vacuum cleaning domain.

\section{Acknowledgements}
This project was supported in part by ONR MURI grant N00014-16-1-2278
from the Office of Naval Research and by NSF IIS grant 1723963.
\bibliography{aaai18-norm-conflict}
\bibliographystyle{aaai}
\end{document}